\documentclass[12pt]{article}
\usepackage{graphics}
\begin{document}

\begin{center}
DYNAMICS OF THE PRIMORDIAL HYDROGEN AND HELIUM (HeI) RECOMBINATION IN THE 
UNIVERSE
\end{center}

\begin{center}
V.K.Dubrovich$^{1,2}$, S.I.Grachev$^{3,}$
\footnotemark[4]\footnotetext[4]{E-mail: stas@astro.spbu.ru}
\end{center}

\begin{center}
$^1${\it Special Astrophysical Observatory, Saint-Petersburg Department,
Saint-Petersburg, Russia}\\
$^2${\it Main Astronomical Observatory of Russian Academy of Sciences}\\
$^3${\it Sobolev Astronomical Institute, Saint-Petersburg State University,
Saint-Petersburg, Russia}\\
\end{center}

\begin{quote}
{\small
{\bf Abstract}. The dependences on $z$
of fractional number densities of H$^+$ and He$^+$ ions are calculated with
a proper allowance for two-photon decays of upper levels of hydrogen and 
parahelium and radiative transfer in intercombination line $2\,^3\!P_1
\leftrightarrow 1\,^1\!S_0$ of helium. It is shown that for hydrogen this
gives corrections for a degree of ionization in no more than a few percents  
but for helium this leads to a significant acceleration of recombination 
compared to the recent papers by Seager et al. (1999; 2000) where these 
effects were ignored.

\medskip

}
\end{quote}

\begin{center}
INTRODUCTION
\end{center}

Dynamics of primordial hydrogen and helium recombination was considered in
a number of papers (see a brief review and references in Seager et al., 2000).
However an actuality of new more accurate and detailed investigations of
this process does not diminish. This is caused by a growing accuracy of new
measurements of the microwave background radiation (CMBR) parameters with the
aim to detect contributions of different new fundamental physical factors -- 
dark matter, dark energy etc.

The most detailed calculations of the matter recombination in the Universe
were fulfilled by Seager et al. (2000) by means of numerical solution of
nonstationary equations for the level populations of hydrogen (300 levels),
HeI (200 levels), HeII (100 levels) and for the number densities of electrons,
protons, hydrogen negative ions H$^-$ and molecules H$_2$ and H$_2^+$ jointly
with an equation for a matter temperature. With all this an average radiation
intensity was taken to be the Planck function at all frequencies except for the
ones in resonant lines for which the Sobolev approximation have been
used. Collisional transitions were also taken into account along with
radiative transitions but their contribution turned out to be negligible (as
was obtained by a number of authors). It should be stressed that for HeI there
were taken into account singlet states (parahelium) as well as triplet states
(ortohelium).

The main result by Seager et al. (2000) concerning HeI recombination consists
in that it goes much more slower than in equilibrium case (according to Saha
equation) and slower than it was obtained by other authors. And as a main
"regulators" of recombination rate appear transitions from the second level
of parahelium -- two-photon ones ($2\,^1\!S_0\leftrightarrow 
1\,^1\!S_0$) and those of in a resonant line $2\,^1\!P_1\leftrightarrow 1\,
^1\!S_0$. It turned out that (as in the case of hydrogen) the results of
multilevel calculations are well described in terms of "effective three level
atom" (offered earlier for hydrogen by Peebles (1968) and by Zeldovich et al. 
(1968)) for parahelium with the usage of a corresponding fitting of the total
recombination coefficient for the upper levels. In this connection Seager et 
al. (1999) created a symplified code recfast.for to compute recombination 
dynamics in terms of "effective three level model" of hydrogen atoms (levels 
$1s$, $2s$, $2p$ $+$ united continuum of upper levels) and helium atoms 
(levels $1s\,^1\!S_0$, $2s\,^1\!S_0$, $2p\,^1\!P_1$ $+$ united continuum of 
upper levels of parahelium).

In the present paper we investigate an influence of additional factors on the
recombination dynamics of hydrogen and helium (HeI), namely -- a contribution 
of two-photon transitions from upper levels of hydrogen and parahelium down to
the ground state and radiative transfer in intercombination line
$2\,^3\!P_1\leftrightarrow 1\,^1\!S_0$ of HeI, which were not taken into 
account explicitely in the code recfast.for by Seager et al. (1999) and 
probably were not taken into account in their multilevel calculations (Seager
et al., 2000). We have considered how one can take into account these factors
in a code recfast.for and have obtaied a significant acceleration of the HeI
recombination. We also have written our own computer code for calculation of
HeI recombination dynamics with a proper allowance for all factors mentioned 
above and have obtained practically the same results as according to a modified
code recfast.for.

\begin{center}
THE MAIN EQUATIONS
\end{center}

Nonstionary equation for a degree of ionization of a chemical element
(hydrogen or helium) $y=N^+/N$ in the uniform expanding Universe can be 
written in a form
\begin{equation}
(1+z)H(z)N(z)\frac{dy}{dz}=\sum_i R_i,
\label{maineq}
\end{equation}
where $N$ is a total number density of atoms and ions of an element, and $N^+$
is a number density of its ions, $H(z)$ is the Hubble factor, $z$ is a 
redshift, $R_i$ is a net rate of transitions from a level $i$ down to the 
ground level 1 defining irreversible recombination. To pass from the time scale
$t$ to the scale of redshifts $z$ we have used an equation $dz/dt=-(1+z)H(z)$.

Equation (\ref{maineq}) holds in assumption that the level populations of
excited states satisfy to the stationary equations of statistical equilibrium
with number densities of electrons and ions available at the moment. This
assumption is really fulfilled because the population of excited states is
defined by permitted transitions (in the field of blackbody background 
radiation) while the number densities of electrons and ions are defined by
much more slower two-photon transitions as well as by a "red-shifting" 
of resonant photons to the longwave region due to the Universe expansion.
We take into account only radiative transitions because as it commonly known
(see e.g. Seager et al., 2000) collisional transitions are negligible. Then
\begin{equation}
R_i=N_i(A_{i1}+B_{i1}J_{1i})-N_1B_{1i}J_{1i},
\label{Reqn}
\end{equation}
where $N_i$ is a population of a level $i$, $A_{i1}$, $B_{i1}$ and $B_{1i}$ are
Einstein coefficients for transitions between level $i$ and the first level, 
$J_{1i}$ is some "average" radiation intensity at the transition frequency. 

Up to the present time the two-photon decay of the second level ($2\,^1\!S_0$ 
for helium) and the "red-shifting" of photons in the main resonace line 
($2\,^1\!P_1\rightarrow 1\,^1\!S_0$ for helium) were considered as the
only regulators of the recombination rate (see e.g. the paper by Seager et al.
(2000) and references therein). However it is clear that similar processes
should be taken into account for the upper levels of hydrogen and parahelium
as well. Moreover, as it will be shown below, the intercombination transition
$2\,^3\!P_1\rightarrow 1\,^1\!S_0$ from the lower state of ortohelium takes a 
noticeable part.

An expression (\ref{Reqn}) for two-photon transitions is rewritten as
\begin{equation}
R_i^{(2q)}=N_iA_{i1}^{(2q)}\left[1-\frac{N_1g_i}{N_ig_1}e^{-h\nu_{1i}/kT}
\right]/(1-e^{-h\nu_{1i}/kT}),
\label{Ri2q}
\end{equation}
and it is supposed that the reverse process is a "capture"
of two photons of a blackbody radiation (with a temperature $T$) with a total 
energy equal to a transition energy $h\nu_{1i}$. Here $\nu_{1i}$ is a 
transition frequency, $g_i$ is a statistical wheight of a level $i$.

Assuming complete frequency redistribution under scatterings and using a
boundary condition that radiation intensity in a shortwave wing of a line
strives to the Planck function we have for resonant transitions
\begin{equation}
R_i=\beta_{1i}N_iA_{i1}\left[1-\frac{N_1g_i}{N_ig_1}e^{-h\nu_{1i}/kT}\right]/
(1-e^{-h\nu_{1i}/kT}),
\label{Ribeta}
\end{equation}
where $\beta_{1k}$ is a probability of a photon escape from a process of
scatterings due to an expansion of the medium (Universe). It is defined 
through the Sobolev optical distance $\tau_{ik}$:
\begin{equation}
\beta_{ik}=(1/\tau_{ik})\left(1-e^{-\tau_{ik}}\right),\quad
1/\tau_{ik}=\frac{4\pi}{hc}\frac{H(z)}{N_iB_{ik}}\left(1-
\frac{N_kg_i}{N_ig_k}\right)^{-1}.
\label{beta}
\end{equation}
Equations (\ref{Ribeta}) and (\ref{beta}) correspond to the Sobolev 
approximation (Sobolev, 1947; see also a review by Grachev (1994), devoted to
some generalizations of this approximation), though within the framework of 
available kynematics (a uniform expansion) it gives an exact solution. 
Substitution of eq. (\ref{beta}) in eq. (\ref{Ribeta}) gives
\begin{equation}
R_i=\frac{8\pi H(z)}{\lambda_{1i}^3}\frac{g_1N_i}{g_iN_1}(1-e^{-\tau_{1i}})
\frac{1-(N_1g_i/N_ig_1)e^{-h\nu_{1i}/kT}}{1-N_ig_1/N_1g_i}/
(1-e^{-h\nu_{1i}/kT}),
\label{Rires}
\end{equation}
where the wavelength of transition $\lambda_{1i}=c/\nu_{1i}$. So far as 
at the begining of He$^+$ recombination (for $z\approx 2700$) we already have
for energies and populations of excited states $h\nu_{1i}\gg kT$ and $N_i\ll 
N_1$ respectively then in eqs. (\ref{Ri2q}) and (\ref{Rires}) in the last
multiplier one can neglect by an exponential term compared to unity and in eq.
(\ref{Rires}) -- also by the second term in the denominator of the fraction.
Then the equations mentioned above are rewritten as
\begin{equation}
R_i^{(2q)}=N_iA_{i1}^{(2q)}[1-(N_1g_i/N_ig_1)e^{-h\nu_{1i}/kT}],
\label{Ri2q1}
\end{equation}
and
\begin{equation}
R_i=\frac{8\pi H(z)}{\lambda_{1i}^3}\frac{g_1N_i}{g_iN_1}(1-e^{-\tau_{1i}})
[1-(N_1g_i/N_ig_1)e^{-h\nu_{1i}/kT}],
\label{Rires1}
\end{equation}
where $N_1=N-N^+$ because an overwhelming part of hydrogen an helium neutral 
atoms are in the ground state at the epoch of recombination.

Thus the problem is reduced to a solution of eq. (\ref{maineq}) with $R_i$ 
at the righthand side according to eqs. (\ref{Ri2q1}) and (\ref{Rires1}) while
the populations of excited states appearing in the righthand sides of these
equations are determined from equations of statistical equilibrium for the
current values of electrons and ions number densities. It should be noted that
such an approach was used by us earlier (Grachev and Dubrovich, 1991) under 
the calculations of hydrogen recombination within the framework of 60-level 
model of atoms. However during an almost all time of helium recombination ($z=
2700 - 1800$) the radiation temperature remains high enough for populations of 
HeI excited states (which are less than 5 eV away from the continuum) to be 
close to equilibrium ones (relative to the continuum) i.e. were defined by 
Boltzmann -- Saha equations with the electron temperature equal to the 
radiation temperature ($T_e=T$):
\begin{equation}
N_i=N_eN^+\frac{g_i}{2g^+g(T_e)}e^{h\nu_{ic}/kT_e},\quad g(T_e)=(2\pi mkT_e)^
{3/2}/h^3,
\label{treq}
\end{equation}
where $N_e$ is an electron number density, $h\nu_{ic}$ is a threshold energy 
of ionization from a level $i$. Then  eqs. (\ref{Ri2q1}) and (\ref{Rires1}) 
are rewritten in the form 
\begin{equation}
R_i^{(2q)}=\frac{N(1-y)}{r_1}g_iA_{i1}^{(2q)}e^{h\nu_{ic}/kT}(1-
r_1e^{-h\nu_{1c}/kT}),
\label{Ri2q2}
\end{equation}
\begin{equation}
R_i=\frac{8\pi H(z)}{\lambda_{1i}^3}\frac{1}{r_1}e^{h\nu_{ic}/kT}
(1-e^{-\tau_{1i}})(1-r_1e^{-h\nu_{1c}/kT}),
\label{Rires2}
\end{equation}
where $r_1=(2g^+/g_1)g(T_e)(1-y)/(N_ey)$.

In view of importance of allowance for two-photon transitions from upper levels
of helium we adduce below a brief derivation of equations for transition 
probabilities according to Dubrovich (1987). The process of simultaneous
emission of two photons by excited atoms is well-known long ago and is described
in text-books (see Berestetskii et al., 1989). Well-known long ago is also the
role of this process in the hydrogen atom for the continuum radiation formation
in the interstellar medium (Kipper, 1950) and in the early Universe (Zeldovich
et al., 1968). However, in these specific cases only one state of hydrogen is
taken into account namely -- $2s$. At the same time, as it was shown by
Dubrovich (1987), for some values of a medium and background radiation
parameters similar decays of upper levels can give noticable and in some cases
a main contribution. Here we will consider this question only in the context of
hydrogen and helium recombination in the early Universe.

From quantum-mechanical selection rules it follows that actually we should
consider $is$ and $id$ states only. An exact expression for a probability
of spontaneous two-photon transition is written according to Berestetskii et 
al. (1989) (eqs. (59.28)) in the form
\begin{equation}
dW=\frac{2^{10}\pi^6\nu^3\nu'^3}{9h^2c^6}\left|\sum_{i',l'}\left[\frac{
(d_\alpha)_{1s,i'l'}(d_\beta)_{i'l',il}}{\nu_{i'l',il}+\nu}+\frac{(d_\beta)_
{1s,i'l'}(d_\alpha)_{i'l',il}}{\nu_{i'l',il}+\nu'}\right]\right|^2d\nu,
\label{dW}
\end{equation}
where $\nu+\nu'=\nu_{il,1s}$, subscripts $il$, $i'l'$ and $1s$ number 
initial, intermediate and final atom states ($i$ is a main quantum number, $l$
is an orbital moment), $(d_{\alpha})_{il,i'l'}$ is a matrix element of a dipole
moment, $\alpha$ and $\beta$ number spatial components of a dipole moment
vector, $\nu$ and $\nu'$ are the frequencies of emitted photons, $h$ is the
Planck constant, $c$ is the velocity of light. Sharp maxima
in eq. (\ref{dW}) under $\nu$ or $\nu'$ $=$ $\nu_{ii'}$ correspond to 
resonances of a cascade transition from an excited level down with an emission
of photons of a discrete spectrum of an atom. Corresponding to them very large
transition probability leads in specific conditions of quasi-equilibrium with a
blackbody radiation to a very high probability of a reverse capture of the same
photons. Strictly speaking this question must be learned more thoroughly 
because "not entirely" resonant transition, but not far from a resonance,
can give contribution under the scheme of "a Lyman quantum escape into the 
wing". In principle this will lead to an additional speeding-up of 
recombination. However a thorough analysis requires a radiative transfer 
solution which we intend to obtain in the next paper. Here we will consider 
only transitions giving continuous distribution of emitted photons i.e. we 
shall assume that $\nu_{i1} -\nu' \sim \nu$. So an obtained rate of 
recombination can be regarded as a lower estimate.

Because we are interested in a two-photon transition to a final state with
a zero orbital moment ($s$ state) then $l$ may be equal only $0$ ($s$ state)
or $2$ ($d$ state) according to selection rules for dipole transitions. For
both cases the value of $l'$ may be equal only $1$ ($p$ state). In this case
an expression for $W$ can be simplified significantly. It can be simplified
still more if to notice that according to summation rule for dipole transitions
[Berestetskii et al., 1989, eq. (52.8)] nearly 90\% of contribution is due to
transitions with $i'=i$ [Berestetskii et al., 1989, eq. (52.6)]. As a result
we come to an expression for a matrix element structure well-known for the
$2s - 1s$ transition in a hydrogen atom. In our case to calculate $W$ we need 
only in a proper allowance for the frequency difference i.e. we must multiply
$A_{2s,1s}$ by $(\nu_{i1}/\nu_{21})^5$ and sum up the two ways of decay (from 
$s$ and $d$ sublevels) with a proper allowance for their statistical wheigts. 
Finally we obtain for hydrogen the following expression:
\begin{equation}
W_{\rm H}\equiv g_iA_{i1}^{(2q)}=54\cdot A_{2s,1s}\cdot[(i-1)/(i+1)]^{2i}
(11i^2-41)/i.
\label{A2qH}
\end{equation}
For large $i$ we have approximately $W_{\rm H} = 89i$ s$^{-1}$. The growth of 
$W_{\rm H}$
with the level number takes place actually up to a some value $i$. 
This is caused by an existence of the limit of the dipole approximation 
applicability. Namely the wavelength of emitted photon (which is in our case
$\sim 2/\nu_{i1}\rightarrow $ const) must be greater than the size of an
excited state orbit ($\sim i^2$). For hydrogen $i_{\rm max} \sim 30$ 
(Beigman and Syrkin, 1983).

Similar consideration can be carry out also for decays of hydrogen-like states
of HeI. For $i > 6-7$ such an approximation is quite rightfull for the matrix
elements $i(s,d) - ip$. It becomes sufficiently accurate for our aims also
for the square of the matrix element $ip\rightarrow 1s$ if we introduce 
correction multiplier 1.15 -- 1.20 which follows from comparing of oscillator
strengths of these transitions for HeI and for hydrogen. And of course 
eq. (\ref{A2qH}) must be renormalized again with a proper allowance for the
frequency of emitted photons i.e. $W$ for hydrogen must be multiplied by
$(24.6/13.6)^5 = 19.4$. Finally we have for $W_{\rm HeI}$:
\begin{equation}
W_{\rm HeI}\equiv g_iA_{i1}^{(2q)}=1045\cdot A\cdot[(i-1)/(i+1)]^{2i}
(11i^2-41)/i. 
\label{A2qHe}
\end{equation}
A limiting condition for $i$ is here the same as for the case of hydrogen.
If dipole approximation is applicable (for $i<40$) then one should take $A=10$
s$^{-1}$ in eq. (\ref{A2qHe}). Otherwise there is an uncertainty connected
with poorly known (both theoretically and experimentally) dependence of $A$ on
a level number. Most probably a contribution of these
levels ($i>40$) is not very large. Approximately it can be taken into account
taking $A=12$ s$^{-1}$. However for precise measurements of the power spectrum
with the aim to obtain information about weak but very important factors an
additional investigation of $A$ is necessary.

An influence of two-photon decays of upper levels on the dynamics of
recombination must be much more significant for HeI than for hydrogen.
This is caused by two circumstances: first -- by much more absolute value of
$W$ and, second, -- by the fact that relation between populations of
$2s$-level and Rydberg levels significantly differs since a ratio of these
populations contains the Boltzmann factor $\exp(-h\nu_{i2}/kT_r)$. 
For hydrogen this factor is equal approximately $3\cdot 10^{-5}$ while for HeI
it is larger approximately at 85 times since an absolute values of energy
differences for HeI are approximately the same but the temperature at which
it recombines is significantly higher. A contribution to a rate of destruction
of "superfluous" Lyman quantums is defined by a product of a population on
a decay probability.

We wrote computer programme to solve eq. (\ref{maineq}) for helium with the
terms in the righthand side of the form (\ref{Ri2q2}) and (\ref{Rires2}) and
along with the two-photon transitions from the second level of parahelium
($i=2s \leftrightarrow 1$) we also took into account transitions from upper
levels ($i=6-40$) which were considered as hydrogen-like (Rydberg ones). For
their energy counted from a threshold of ionization we take $h\nu_{ic}\approx 
1\,{\rm Ry}/i^2$ and for Einstein coefficients we use eq. (\ref{A2qHe}).
Moreover along with the resonant transition from the second level of
parahelium $i=2p\,^1\!P_1\leftrightarrow 1s\,^1\!S_0$ (thereafter we wright
for the sake of simplicity $i=2p\leftrightarrow 1$) it was also taken into 
account spin-forbidden one-photon transition from the second level of
ortohelium: $2p\,^3\!P_1\leftrightarrow 1s\,^1\!S_0$ for which we use the value
of Einstein coefficient $A_{2\,^3\!P_1,1\,^1\!S_0}=233$ s$^{-1}$ according to 
Lin et al., 
1977. It should be stressed that for the degree of ionization of hydrogen which
is contained in equation $N_e=N_{\rm H}^++N_{\rm He}^+$ we have used Saha
equation (equilibrium ionization) which holds for sufficiently large $z$ where
the main HeI recombination takes place.

We also introduced corresponding additions into the programme recfast.for by
Seager et al. (1999) in which two-photon transitions were taken into account
only from the second level ($i=2s\leftrightarrow 1$) and Einstein 
coefficients $A_{2s,1}^{(2q)}$ for hydrogen and parahelium ($\Lambda_{\rm H}$ 
and $\Lambda_{\rm HeI}$ in notations by Seager et al., 1999) $\Lambda_{\rm H}=
8.22$ s$^{-1}$ and $\Lambda_{\rm HeI}=51.3$ s$^{-1}$. To take into account
two-photon transitions from upper levels it needs evidently according to eq.
(\ref{Ri2q2}) to make a replacement
\begin{equation}
\Lambda\rightarrow \Lambda+\sum_{i=i_0}^{i_N}g_iA_{i1}^{(2q)}
e^{h(\nu_{ic}-\nu_{2s,c})/kT}.
\label{Lamb}
\end{equation}
Further, in the programme recfast.for a contribution of transition 
$i=2p\leftrightarrow 1$ for hydrogen and parahelium is described by the
factors $K_{\rm H}$ and $K_{\rm HeI}$ respectively where $K=\lambda_{1,2p}/
[8\pi H(z)]$ so that a contribution of other lines (for one-photon transitions)
can be taken into account (in accordance with eq. (\ref{Rires2})) by the
following replacement:
\begin{equation}
\frac{1}{K}\rightarrow \frac{1}{K}\left[1+\sum_i
(\lambda_{1,2p}/\lambda_{1i})^3e^{h(\nu_{ic}-\nu_{2p,c})/kT}(1-e^{-\tau_{1i}})/
(1-e^{-\tau_{1,2p}})\right],
\label{KHe}
\end{equation}
and for the main resonant transition during the whole recombination an
optical depth $\tau_{1,2p}\gg 1$ so that the corresponding exponential term
can be omitted.
  
As the computations have shown the results obtained under our own programme
and under the programme recfast.for modified as pointed above are
practically coincide (for the values of parameters of the Universe model
accepted now).

\begin{center}
RESULTS OF COMPUTATIONS
\end{center}

Parameters in the problem are equilibrium temperature of the microwave
background $T_0$, Hubble factor $H_0$, the ratio of the total density to the
critical one $\Omega_{\rm total}$, the rati® of the baryon density to the 
critical one $\Omega_{\rm B}$, the ratio of the dark matter to the critical 
one $\Omega_{\rm DM}$, the ratio of the density caused by the $\Lambda$-term to
the critical one $\Omega_{\Lambda}$ at the present day epoch and the content of
the primordial helium (by mass) $Y$. Test computations were fulfilled for
$T_0=2.728$ K, $\Omega_{\rm total}=1$, $Y=0.24$,
$H_0=70$ (km/s)/Mpc, $\Omega_{\rm B}=0.04$, $\Omega_{\rm DM}=0.26$ and
$\Omega_{\Lambda}=0.7$.  

\begin{figure}[t]
\vspace*{-4.5cm}
\centering
\resizebox{1.0\textwidth}{!}{\includegraphics{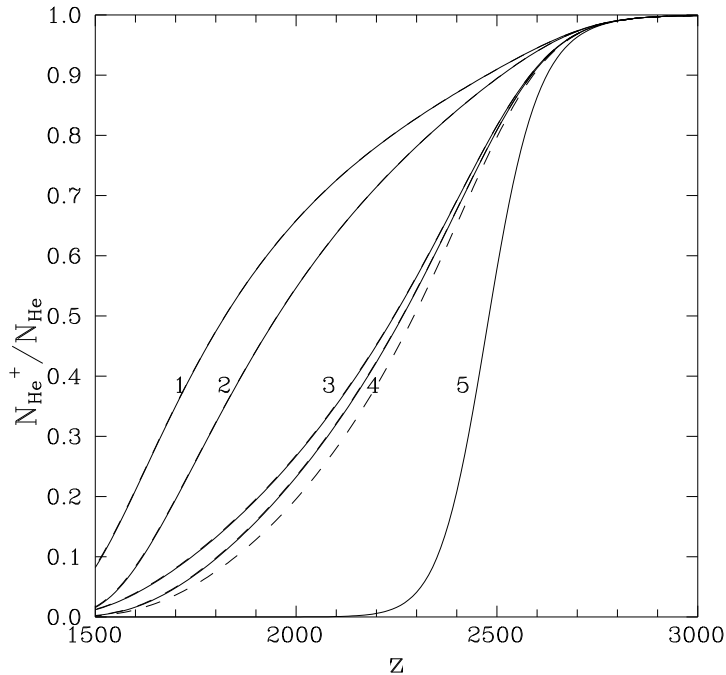}}
\vspace{-11cm}
\caption{The profiles of helium ionization degree. Numbers near the curves:
1 -- only "main" transitions $2s\leftrightarrow 1$ and $2p\leftrightarrow 
1$ of parahelium, 2 -- main $+$ transition in a line of ortohelium, 3 -- main 
$+$ transitions from upper levels of parahelium ($A=10$ s$^{-1}$), 4 -- main 
$+$ both additional ways (continuous line -- $A=10$ s$^{-1}$, broken line -- 
$A=12$ s$^{-1}$), 5 -- 
equilibrium case (Saha equation). Values of parameters: $\Omega_{\rm total}=
1$, $T_0=2.728$ K, $H_0=70$ (km/s)/Mpc, $\Omega_{\rm B}=0.04$, $\Omega_{\rm 
DM}=0.26$, $\Omega_{\Lambda}=0.7$. Helium content (by mass) $Y=0.24$.}
\end{figure}

Fig. 1 shows the results of computations. Numbers 1, 2, 3 and 4 correspond to
a succesive allowance for different ways of irreversible helium recombination:
1 -- only "main" transitions $2s\leftrightarrow 1$ and $2p\leftrightarrow 
1$ of parahelium, 2 -- main $+$ transition in ortohelium line, 3 -- main $+$
transitions from the upper levels ($i=6 - 40$) of parahelium for $A=10$ s$^{-1}$
in eq. (\ref{A2qHe}), 4 -- main $+$ both additional ways (continuous line -- 
for $A=10$ s$^{-1}$ in eq. (\ref{A2qHe}), broken line -- for $A=12$ s$^{-1}$, 
5 -- equilibrium case
(Saha equation). One can see from the Fig. 1 that succesive allowance for
additional ways of irreversible recombination significantly enhances the rate
of HeI recombination. At the same time our results practically coincide
(undistinguished on the Figure) with the ones obtained under the programme
recfast.for modified by the way stated in the preceeding section.

With the aid of the programme recfast.for we took into account additional ways 
of irreversible recombination for hydrogen too. These ways are two-photon
transitions from the levels $i=3 - 40$ down to the first level and radiative
transfer in the corresponding Lyman lines. The first way was taken into
account according to eqs. (\ref{Lamb}) and (\ref{A2qH}) and the second -- 
according to eq. (\ref{KHe}) while exponents in round brackets in this equation
were omitted because the Sobolev thicknesses in Lyman lines $\tau_{1i}\gg 1$. 
It turned out that the first way leads to decrease of hydrogen ionization
degree by no more than 4.2\% and the second -- by no more than 1\% and 
in sum -- by no more than 4.3\% (for the same values of the model parameters as
in the case of helium). Results are in the Fig. 2. It should be stressed
that we have fulfilled in addition similar computations using our own programme
(Grachev and Dubrovich, 1991) based on solution of eq. (\ref{maineq}) jointly
with equations of statistical equilibrium for the level populations of hydrogen
atoms within the framework of 60-level model of an atom and we have obtained 
practically the same result as in the Fig. 2.

\begin{figure}[t]
\vspace*{-4.5cm}
\centering
\resizebox{1.0\textwidth}{!}{\includegraphics{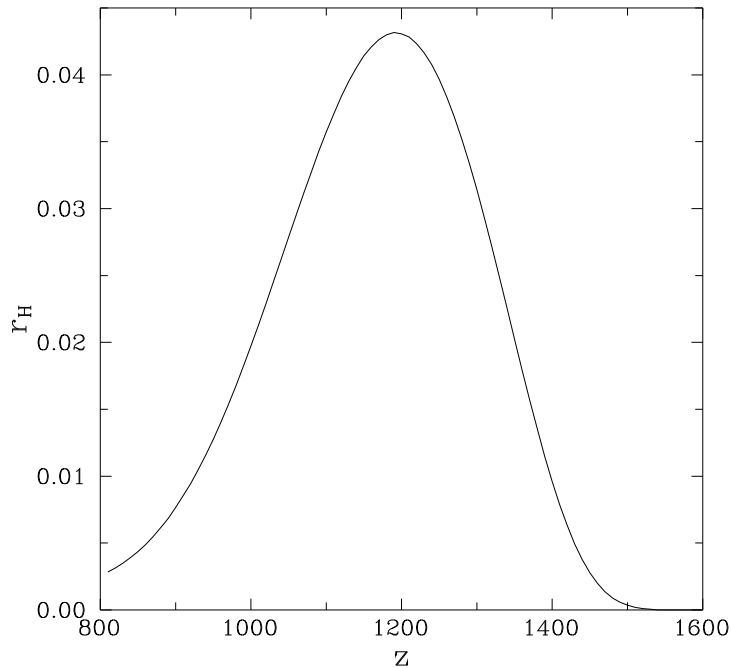}}
\vspace{-11cm}
\caption{An influence of additional transitions on the profile of hydrogen
ionization degree $y_{\rm H}=N_{\rm H}^+/N_{\rm H}$. On the ordinate axis is 
$r_{\rm H}=2(y_{\rm H}^{\rm old}-y_{\rm H}^{\rm new})/(y_{\rm H}^{\rm old}+
y_{\rm H}^{\rm new})$, where $y_{\rm H}^{\rm old}$ corresponds to  
"main" transitions $2s\leftrightarrow 1$ and $2p\leftrightarrow 
1$ only, and $y_{\rm H}^{\rm new}$ corresponds to main $+$ transitions from 
upper levels. Values of parameters are the same as for the Fig. 1.}
\end{figure}

\begin{center}
CONCLUSIONS
\end{center}

Computations of hydrogen and HeI recombination dynamics are made
with a proper allowance for two-photon decays of upper levels of hydrogen
and parahelium and radiative transfer in intercombination line $2\,^3\!P_1
\leftrightarrow 1\,^1\!S_0$ of helium. It is shown that this leads to
changes of hydrogen recombination rate accessible to discover under the
programme "PLANCK". Obtained results are important for a correct evaluation
of small factors defined by an existence of dark matter, baryon and nonbaryon 
parts of the substance mass in the Universe, relativistic (at the moment of
recombination) particles -- neutrino and possibly axions and other low-mass
low-interacting particles. It is shown that an allowance for new factors of 
destruction of supraequilibrium Lyman photons significantly accelerates HeI
recombination compared to predictions according to the papers by Seager et al.
(1999; 2000) where these effects were not taken into account. This is also
important for the correct estimation of the helium role in the hydrogen
recombination. Moreover it is very important for determination of profiles
and intensities of hydrogen and HeI recombination lines. In particular an
important role of high levels in irreversible recombination leads to an
appearance of absorption lines in the CMBR spectrum caused by transitions in
Balmer lines of hydrogen and in different more complex lines of helium.
This moment is principally important from the point of view of detection and
identification of such lines. As a next step in this direction it is necessary
to consider a role of weakly nonresonant transitions from upper levels which
can lead to an additional acceleration of recombination for hydrogen as well as
for helium. 

As concerns an influence of our refinements of HeI recombination dynamics on
the theoretical power spectra of microwave background radiation it can be 
estimated very roughly from Fig. 17 in the paper by Seager et al. (2000) where
their results are compared with the results by Hu et al. (1995) obtained
actually assuming equilibrium HeI ionization (according to Saha equation).
So, since our curve of variation of HeI ionization degree lies practically "in 
the middle" between equilibrium one (according to Saha equation) and that from
Seager et al. (see Fig. 1) then one can expect that with our refinements the
discrepancy with the results by Hu et al. will become 2 times smaller (e.g.
for multipole $l=1500$ it will be 1\% instead of 2\%). At the same time an
uncertainty of $A$ (see Fig. 1, curves 4) estimated approximatly at 20\% should
give the same relative uncertainty for correction of theoretical power 
spectrum.

This work is supported by RFBR grant \# 02-02-16535, by the grant of the
President of Russian Federation for supporting of Leading scientific schools
(grant NSh--1088.2003.2), through the federal programme "Astronomy" and by
Goscontract 40.022.1.1.1106.

\begin{center}
REFERENCES
\end{center}

%\begin{thebibliography}{99}
\begin{enumerate}
\item
Beigman I.L., Syrkin M.I., Preprint \# 295 (Moskow.: FIAN, 1983) [in Russian].
\item
Berestetskii V.B., Lifshitz E.M., Pitaevskii L.P., {\it Quantum 
Electrodynamics} (Moskow: Nauka, 1989) [in Russian].
\item 
Dubrovich V.K., Optics and Spectroscopy. {\bf 63}, 439 (1987) [in Russian].
\item
Grachev S.I., Transac. Astron. observ. Saint-Petersburg State Univ. {\bf 44}, 203
(1994) [in Russian].
\item
Grachev S.I., Dubrovich V.K., Astrophysics. {\bf 34}, 249 (1991) [in Russian].
\item
Hu W., Scott D., Sugiyama N., White M., Phys. Rev. D. {\bf 52}, 5498 (1995).
\item
Kipper A.Ya., in. {\it On the development of soviet science in ESSR, 1940-1950}
(Tallinn, 1950) [in Russian].
\item
Lin C.D., Johnson W.R., Dalgarno A., Phys. Rev. A. {\bf 15}, 154 
(1977).
\item
Peebles P.J.E., Astrophys. J. {\bf 153}, 1 (1968).
\item
Seager S., Sasselov D.D., Scott D., Astrophys. J. {\bf 523}, L1 (1999).
\item
Seager S., Sasselov D.D., Scott D., Astrophys. J. Suppl. Ser. 
{\bf 128}, 407 (2000).
\item
Sobolev V.V., {\it Moving envelopes of stars} (Leningrad: Leningrad State Univ.
Press, 1947 [in Russian]; English transl., Cambridge: Harvard Univ. Press,
1960).
\item
Zeldovich Ya.B., Kurt V.G., Sunyaev R.A., Zh. Exsper. Teor. Fiz. {\bf 55}, 
278 (1968) [in Russian]; English transl., Soviet Phys.--JETP Lett., {\bf 28},
146 (1969).
\end{enumerate}

\end{document}